\documentclass[a4paper,12pt]{article}
\usepackage{amsmath,amssymb,array,float,graphics}
\usepackage{pstricks,pst-node,epsfig}
\usepackage[figuresleft]{rotating}
\usepackage{rotfloat}
\usepackage{cite}
\numberwithin{equation}{section}

\newcommand{\ii}{\mathrm{i}}

\newcommand{\dd}{\mathrm{d}}
\newcommand{\pd}{\partial}

\newcommand{\e}{\mathrm{e}}
\newcommand{\ket}[1]{\left|#1\right\rangle}
\newcommand{\bra}[1]{\left\langle #1\right|}
\newcommand{\bracket}[2]{\left\langle%
#1\left.\right|#2\right\rangle}

\newcommand{\tr}{\mathop{\mathrm{Tr}}\nolimits}

\newcommand{\R}{\mathbb{R}}

\newcommand{\I}{\mathbb{I}}

\newcommand{\ft}[2]{{\textstyle\frac{#1}{#2}}}
\def\tilde{\widetilde}
\def\1bar{1\hskip -.275cm -}
\def\2bar{2\hskip -.275cm -}
\def\3bar{3\hskip -.275cm -}

\newsavebox{\uuunit}

\newcommand{\nn}{\nonumber}

\newcommand{\nc}{\newcommand}
\nc{\la}{\lambda} \nc{\alf}{\alpha} \nc{\tht}{\theta}
\nc{\eps}{\epsilon} \nc{\ga}{\gamma} \nc{\Ga}{\Gamma}
\nc{\De}{\Delta} \nc{\de}{\delta} \nc{\si}{\sigma}
\nc{\ka}{\kappa} \nc{\om}{\omega} \nc{\qq}{\quad\quad}
\nc{\nf}{\infty} \nc{\dl}{\mathop{\smash{\cal L}}}
\nc{\ol}{\overline} \nc{\beq}{\begin{equation}}
\nc{\barr}{\begin{array}} \nc{\earr}{\end{array}}
\nc{\eeq}{\end{equation}} \nc{\beqa}{\begin{eqnarray}}
\nc{\dst}{\displaystyle}\nc{\pt}{\partial}
\nc{\eeqa}{\end{eqnarray}} \nc{\nnb}{\nonumber}
\nc{\bs}{\backslash}        \nc{\mbb}{\mathbb}
\nc{\brm}{\begin{remunerate}} \nc{\erm}{\end{remunerate}}
\nc{\vareps}{\varepsilon} \nc{\tb}{\tilde\beta_0} \nc{\ts}{\tilde
s} \nc{\tth}{\tilde \theta}
\newcounter{muni}
\usecounter{muni}
  \nc{\lapdec}{\mathop{\Delta}}

\newenvironment{remunerate}{\begin{list}{{\rm \arabic{muni}.}}
{\usecounter{muni}
\setlength{\leftmargin}{0pt}\setlength{\itemindent}{38pt}}}{\end{list}}
\nc{\cre}{\color[rgb]{1.00,0.00,0.00}}
\nc{\cgr}{\color[rgb]{0.00,1.00,0.00}}

\makeatletter
\def\hlinewd#1{%
\noalign{\ifnum0=`}\fi\hrule \@height #1 %
\futurelet\reserved@a\@xhline} \makeatother

\begin{document}
\title{SL(2) spin chain and spinning strings on $AdS_5\times S^5$}

\author{S. Bellucci, P.-Y. Casteill, J.F. Morales$^\dagger$, C. Sochichiu\thanks
{On leave from~: Bogoliubov Lab. Theor. Phys., JINR, 141980 Dubna, Moscow Reg.,
RUSSIA and Institutul de Fizic\u a Aplicat\u a A\c S, str. Academiei, nr. 5,
Chi\c{s}in\u{a}u, MD2028
MOLDOVA.}\\
{\it INFN -- Laboratori Nazionali di Frascati,}\\
{\it Via E. Fermi 40, 00044 Frascati, Italy}\\
$^\dagger$ {\it Dipartimento di Fisica Teorica, Universita di
Torino}.}

\maketitle
\begin{abstract}
We derive the coherent state representation of the integrable spin
chain Hamiltonian with symmetry group SL(2,$\R$). By passing to
the continuum limit, we find a spin chain sigma model describing a
string moving on the hyperboloid SL(2,$\R$)/U(1). The same sigma
model is found by considering strings rotating with large angular
momentum in $AdS_5\times S^5$. The spinning strings are identified
with semiclassical coherent states built out of SL(2,$\R$) spin
chain states.
\end{abstract}

\tableofcontents

\section{Introduction}

AdS/CFT correspondence
\cite{Maldacena:1998re,Gubser:1998bc,Witten:1998qj} relates string
theories on AdS spaces and gauge theories living on the AdS
boundaries. The typical example is the correspondence between
${\cal N}=4$ SYM and IIB string on $AdS_5\times S^5$.
 The duality maps bulk gravity
 and string fields to excitations of the gauge theory living
 in the holographic screen and puts in correspondence their dynamics.
 The two dual descriptions  are
 complementary in the sense that, when spacetime curvatures are large and
 a low energy supergravity description is reliable, the gauge coupling constant
 is large and viceversa, when the gauge coupling is small, the supergravity
  description breaks down and one has to consider the full string theory on AdS.
 Unfortunately, string theory on AdS, like gauge theories beyond the perturbative
 regime, is hard to deal with making most tests of the correspondence out of reach.
 In supersymmetric models which are
dealt with, however, one can test the correspondence using BPS
states, i.e. states that preserve part of supersymmetry and whose
properties are protected by it against quantum corrections, even
strong ones (see \cite{Aharony:1999ti} for a review and a complete
list of references).
 More recently, strings on AdS have been tested against
 free SYM
\cite{
Bianchi:2003wx,Beisert:2003te, Gopakumar:2003ns,Aharony:2003sx,
Gopakumar:2004qb,Beisert:2004di,Bonelli:2004ve}, but how to bring
interactions into the game is far from clear.

In \cite{Berenstein:2002jq}, the authors  considered the
holographic correspondence near null geodesics of $AdS_5\times
S^5$, where the geometry looks like a pp-wave \cite{Blau:2001ne}.
 On the gauge theory side, this corresponds to consider SYM operators
 with large ${\cal R}$-symmetry charge $J$.
String theory on a pp-wave is known  to be integrable
\cite{Metsaev:2001bj,Metsaev:2002re}, allowing for quantitative
tests of the correspondence beyond the supergravity level.
  This appears possible since corrections to the
string theory for states not far from BPS can be interpreted as
``semiclassical'' \cite{Gubser:2002tv,Frolov:2002av}. Multispin
solutions were further found and analyzed in
\cite{Tseytlin:2003ac,Frolov:2003xy,Frolov:2003qc,Frolov:2003tu,
Arutyunov:2003uj,Beisert:2003ea,Beisert:2003xu}. In this approach,
energies of classical string solutions are compared to anomalous
dimensions of SYM operators with large $J$ charges (see
\cite{Tseytlin:2003ii} for a review and a complete list of
references).

On the other hand there has been an enormous progress in the
understanding of ${\cal N}=4$ SYM dynamics. In a series of nice
papers \cite{Minahan:2002ve,Beisert:2003yb, Beisert:2003jj} the
planar limit of the dilatation operator for ${\cal N}=4$ SYM was
identified with the Hamiltonian of integrable spin chains, while
non planar contributions  were realized in
\cite{Beisert:2002ff,Bellucci:2004ru,Bellucci:2004qx} in terms of
a joining-splitting spin chain operator mimicking string
interactions.

  There are two representative sectors of the SYM theory~: the su(2) sector
 associated to scalar impurities (excitations along $S^5$) and the sl(2)
 sector associated to vector impurities (excitations along $AdS_5$).
 In \cite{Kruczenski:2003gt}
 the su(2) sector was explored using a spin chain sigma model description,
 and classical string spinning on $S^5$ were identified
with semiclassical coherent states built out of su(2) eigenstates
in the spin chain system. This was inspired by the observation
that in the thermodynamic limit the spin chain Hamiltonian and
higher conserved charges fit those coming from semiclassical
string sigma model on $AdS_5\times S^5$
\cite{Arutyunov:2003rg,Arutyunov:2004xy}.
 An extensive check of
the correspondence beyond one loop \cite{Kruczenski:2004kw} and
the generalization to the compact su(3), so(6) sectors was worked
out in
\cite{Hernandez:2004uw,Kristjansen:2004za,Kruczenski:2004cn}.

The aim of this paper is to test the
 non-compact sector of the
theory  and consider the sl(2) case\footnote{In this paper sl(2)
will always refers to the real form sl(2,$\R$).}. It corresponds
to SYM operators made out of a single scalar and its derivatives
along a fixed direction. States in this sector carry charges
associated to both the ${\cal R}$-symmetry SO(6) and conformal
SO(4,2) groups. Accordingly sl(2) spin states corresponds to
strings spinning in both $S^5$ and $AdS_5$. The preliminary task
will be, as in \cite{Kruczenski:2003gt}, to derive a sigma model
representation of the spin chain dynamics. The su(2) case was well
studied, in part because of important applications to condensed
matter \cite{Fradkin:book}. Unlike the su(2) case, spin chain and,
respectively, sigma models with non-compact symmetry groups are
less familiar (as far as we know) to the condensed matter
literature. Here we combine the coherent state representation of
the sl(2) algebra and the spin chain Hamiltonian in order to
derive a path integral formulation of the gauge dynamics. In the
continuum limit this Hamiltonian results into a sigma model
describing a string moving on an hyperboloid. sl(2) spin
chain/string sigma models were considered earlier in
 \cite{Stefanski:2004cw}.

The plan of the paper is as follows. In Section 2 we review the
representations and the construction of coherent states of the
sl(2) algebra. In Section 3 we derive a coherent state
representation of the sl(2) spin chain hamiltonian and a path
integral formulation of its dynamics. By passing to the continuum
limit we found a two-dimensional sigma model on an hyperboloid.
 In Section 4 we describe the string duals and match the spin chain sigma model
 to that following from a string spinning fast on $S_\phi^1\times S_\varphi^1$ with
 $S_\phi^1\in$~$AdS_5$ and $S_\varphi^1\in S^5$. In Section 5 we draw some conclusions.

\section{sl(2) coherent states}

In this section we collect the main properties of sl(2) coherent
states. We refer the reader to \cite{book:perelomov} for details.

\subsection{sl(2) and its representations}

We start by describing the sl(2) algebra and its representations.
The algebra sl(2) $\sim$ su(1,1) $\sim$ so(2,1) $\sim$ sp(2,R) is
non-compact and it is defined by the commutation relations
\begin{equation}\label{sl2-comm}
  \left\{\begin{array}{ll}
\left[J_-,J_+\right]&=2J_0,\\
\left[J_0,J_\pm\right]&=\pm J_\pm.
\end{array}\right.
\end{equation}
The operators $J_\pm$ are related to generators $J_{1,2}$, $i=1,2$
through
\begin{equation}
  J_{\pm}=J_2\mp \ii J_1,
\end{equation}
with the following commutation relations:
\begin{equation}
  [J_0,J_1]=\ii J_2,\quad[J_1,J_2]=-\ii J_0,\quad [J_2,J_0]=\ii J_1~ .
\end{equation}
 It is not difficult to verify that the quadratic operator
\begin{equation}\label{casimir}
  \hat{C}_2=J_0^2-J_1^2-J_2^2=J_0^2-\frac{1}{2}(J_+J_-+J_-J_+)
\end{equation}
commutes with all generators, i.e. it is a Casimir of the algebra.
Therefore, it is proportional to unit matrix
\[
  \hat{C}_2=j(j-1)\I.
\]
The number $j$ labels the irreducible representations.
 The algebra $SU(1,1)$ is noncompact, so, unlike the case of
 $SU(2)$, all its unitary irreducible representations are
 infinite-dimensional. The irreducible representations fall into three types of
 series: principal, discrete and supplementary.
 For the
discrete series of representations of sl(2), the spin $j$ takes
the values $j=1,3/2,2,5/2,\dots$. For the universal covering group
$\tilde{\mathrm{sl}}$(2) this is extended to arbitrary positive
numbers $0<j<\infty$. The expressions in  this section will always
refer to the later type of representations. For applications to
SYM gauge theories we will later specify to the case $j\to \ft12$.

 The states in a spin $j$ representation of $SU(1,1)$ will be denoted
by $\ket{j,j+m}$ with $m=0,1,2,\ldots$ The unitarity of the
representation implies that one can choose an orthonormal basis
for the spin states:
\[
  \bracket{j,m}{j,m'}=\delta_{m,m'}.
\]
The representations for the discrete series of sl(2) are
essentially the analytic continuation of those of su(2). The
action of the generators on the states of spin $m$ of
representation $j$ is given by\footnote{Alternatively the algebra
can be represented in the space of functions $f(z)$ holomorphic
within the unit circle (generalized Fock--Bargmann representation)
by $J_+=z^2\pd_z+2jz$, $J_-=\pd_z$, $J_0=z\pd_z+j$.}
\begin{align}\label{action:gen}
  J_+\ket{j,j+m}&=\sqrt{(m+1)(m+2j)}\ket{j,j+m+1},\\
  J_-\ket{j,j+m}&=\sqrt{m(m+2j-1)}\ket{j,j+m-1},\\
  J_0\ket{j,j+m}&=(m+j)\ket{j,j+m}.
\end{align}
In contrast to su(2) representations the action of $J_+$ never
ends and therefore the representation is infinite dimensional. We
will mainly concern ourselves with the spin $j=1/2$ case
\begin{align}\label{action:one-half}
  J_+\ket{m}&=(m+1)\ket{m+1},\\
  J_-\ket{m}&=m\ket{m-1},\\
  J_0\ket{m}&=(m+\ft12)\ket{m}.
\end{align}
with $\ket{m}\equiv \ket{\ft12,\ft12+m}$ from now on.

\subsection{Coherent states} (Generalized) coherent states in the
Perelomov's sense are given by the action of a finite group
transformation on a particular state $\ket{\psi_0}$. It is
suitable to take this state to be the vacuum, i.e. the lowest
weight state $\ket{0}=\ket{j,j}$ of a spin $j$ representation of
sl(2). The coherent state will be parameterized by a point
\[
  \vec{n}=(\cosh\rho,\sinh\rho\sin\phi,\sinh\rho\cos\phi)
\]
on the upper sheet of two-sheet hyperboloid:
\[
\vec{n}^2=  n_0^2-n_1^2-n_2^2=1,\qquad n_0>0~.
\]
We choose the following definition for the coherent state:
\begin{equation}\label{coh-st:gen}
  \ket{\vec{n}}=D(\vec{n})\ket{0}=\e^{\xi J_+- \bar{\xi} J_-}\, \ket{0} \qquad\qquad
  \xi=\ft12\, \rho\, \e^{\ii \phi}~.
\end{equation}
The element $D(\vec{n})$ is the matrix of the hyperbolic rotation
which maps the ``north pole'' $\vec{n}_0=(1,0,0)$ into the point
$\vec{n}$ on the hyperboloid (see Figure \ref{fig:mafigure2}
below). The operators $D(\vec{n})$ do not form a group, but the
multiplication law is $
D(\vec{n}_1)D(\vec{n}_2)=D(\vec{n}_3)\e^{\ii\Phi(\vec{n}_1,\vec{n}_2)\,J_0}
$ where $\Phi(\vec{n}_1,\vec{n}_2)$ is the area of the hyperbolic
triangle with vertices $\vec{n}_1,\vec{n}_2$ and
$\vec{n}_0=(1,0,0)$.
\\[0.5cm]
Next we list the main properties of the coherent states
$\ket{\vec{n}}$.
\\
{\bf 1)} Coherent state content:\nopagebreak

Using \eqref{action:gen}, coherent states can be expanded in
 the spin state basis $\ket{j,j+m}$ building the sl(2)
 representation. The expansion coefficients are given in
 \cite{book:perelomov}:
\begin{equation}\label{cm}
  \ket{\vec{n}}=\sum_{m=0}^{\infty}c_{j,m}\ket{j,j+m},\qquad
  c_{j,m}=\frac{1}{\cosh(\ft{\rho}{2})^{2j}}
\left(\frac{\Gamma(m+2j)}{m!\,\Gamma(2j)}\right)^{\frac12}\,\e^{\ii\,m\,\phi}\,\tanh(\ft{\rho}{2})^{m}.
\end{equation}
The expansion is such that
$$\bra{\vec{n}\, }\vec{n}\rangle=1~.$$
 We are mostly interested in the particular case $j=\ft12$. The expansion
 coefficients drastically simplifies in this case
\begin{equation}\label{cm2}
  \ket{\vec{n}}=\frac{1}{\cosh(\ft{\rho}{2})}\sum_{m=0}^{\infty}
\e^{\ii\,m\,\phi}\, \tanh(\ft{\rho}{2})^{m}\,\ket{m}~,
\end{equation}
 where as before $\ket{m}\equiv \ket{\ft12,\ft12+m}$.
\\\\
{\bf 2)} Coherent states are over-complete.
\\
 Using (\ref{cm}) is not difficult to show that the unity
operator $\I$ can be
 written in terms of the coherent state $\ket{\vec{n}}$ as
\begin{equation}\label{res-un}
  \I=\frac{2j-1}{4\pi}\int\sinh \rho\,\dd \rho\,\dd \phi
  \ket{\vec{n}}\bra{\vec{n}}
  \equiv
 \int\dd^2n\ket{\vec{n}}\bra{\vec{n}} .
\end{equation}
Notice that in this formula the point $j=\ft12$ is singular but
the limit $j\to \ft12$ is well defined (see appendix \ref{acs} for
details). In the rest this limit is always assumed.
\\\\
{\bf 3)} Coherent states are not orthonormal
\\
\begin{equation}\label{scalar} \bra{\vec{n}_{1} }\vec{n}_2\rangle
=\left(\cosh(\frac{\rho_{1}}{2})
\cosh(\frac{\rho_{2}}{2})-\sinh(\frac{\rho_{1}}{2})
\sinh(\frac{\rho_{2}}{2})\,\e^{\ii\,(\phi_2-\phi_1)}
\right)^{-2j}\, .
\end{equation}
\\
{\bf 4)} To each coherent state $\ket{\vec{n}}$ we can associate a
point $\vec{n}$ on the
 hyperboloid via the important property
\\
$$ \bra{\vec{n} }\, \vec{J}\, \ket{\vec{n} } =j\, \vec{n}~.$$
 This property is demonstrated in appendix \ref{acs}.
 In particular, we notice that
$$ \bra{\vec{n} }\, J_0\, \ket{\vec{n} } =j\,\cosh\rho~.$$
This means that $\cosh\rho$ measures the ``average spin'' of the
coherent state.

\section{Hamiltonian in the coherent state basis}

In this section we derive a coherent state representation of the
sl(2) spin chain Hamiltonian and a path integral formulation of
the spin chain dynamics. We follow closely \cite{Fradkin:book}
where the derivation of the su(2) ferromagnetic sigma model is
presented in full details.
 Unlike the more familiar su(2) case, sl(2) is a non-compact algebra
 and its representations are
 infinite-dimensional. This will result into a non-polynomial
 Hamiltonian.

 Here we consider the spin chain describing
  one-loop planar anomalous dimensions for SYM operators in the sl(2) sector.
 The Hamiltonian is given by
  \cite{Beisert:2003jj}
$$
  H=\sum_{k=1}^{L}H_{k\,k+1}
$$
with
\begin{eqnarray} H_{k_1 k_2}\ket{m_{k_1} m_{k_2}} &=&
   \left[h(m_{k_1})+h(m_{k_2})\right]\ket{m_{k_1} m_{k_2}}\nn\\
  &&-\sum_{l=1}^{m_{k_1}}\frac{1}{l}\ket{m_{k_1}-l,m_{k_2}+l}
  -\sum_{l=1}^{m_{k_2}}\frac{1}{l}\ket{m_{k_1}+l,m_{k_2}-l}
  \label{Ham:pair}
\end{eqnarray} and
$$
  \ket{m_{k_1} m_{k_2}}\equiv\ket{m_{k_1}}\otimes\ket{m_{k_1}}~~,
\qquad h(m)=\sum_{i=1}^m\frac1i~~~~.
$$
The underscripts $k_{i}=1,2,\ldots L$ specify the spin chain site
where the corresponding state $\ket{m_{k_i}}$ leaves.

The Hamiltonian (\ref{Ham:pair}) has a nice representation in the
coherent state basis $\ket{\vec{n}}$. Indeed, using (\ref{cm2})
one can compute the average of $H_{k\, k+1}$ over two-site
coherent states $\ket{\vec{n}_k \vec{n}_{k+1}} \equiv
\ket{\vec{n}_k}\otimes \ket{\vec{n}_{k+1}}$. This was first done
in \cite{Stefanski:2004cw} (see appendix \ref{aham} for an
independent derivation). The result reads:
\begin{eqnarray}\label{Ham:final}
   \bra{\vec{n}_k \vec{n}_{k+1}} \, H_{k\,k+1}\, \ket{\vec{n}_k \vec{n}_{k+1}}
  &=&  \log\left(\frac{1+\vec{n}_k.\vec{n}_{k+1}}{2}\right)\nn\\
 &
 =&\log\left(1-\frac{\left(\vec{n}_k-\vec{n}_{k+1}\right)^2}{4}\right)~.
\end{eqnarray}
Here and below the dot product is defined via
$$\vec{n}\cdot\vec{m}=n_0 m_0-n_1 m_1-n_2 m_2~.$$

 In order to pass to path integral representation, one considers the
Hamiltonian as the generator of time translations in the spin
chain system and discretizes the time interval $\Delta T$.
 At a given time $t_s$ we introduce the ``spin
field vector" \begin{equation} \ket{\vec{n}_s}\equiv
\ket{\vec{n}_{1,s}}\otimes \ket{\vec{n}_{2,s}}\otimes \ldots
\otimes\ket{\vec{n}_{L,s}} \end{equation} with
$\vec{n}_{k,s}\equiv \vec{n}_k(t_s)$ specifying the spin chain
configuration at the $k^{\rm th}$ site and time $t_s$. Each
$\vec{n}_{k}$ describes therefore a trajectory on the hyperboloid.
Alternatively, the ``$L$-plet" $\{ \vec{n}_k(t_s) \}$ specifies
the spin chain profile at a fixed time $t_s$.

\begin{figure}[h]
            \begin{center}
               \scalebox{0.4}{\input{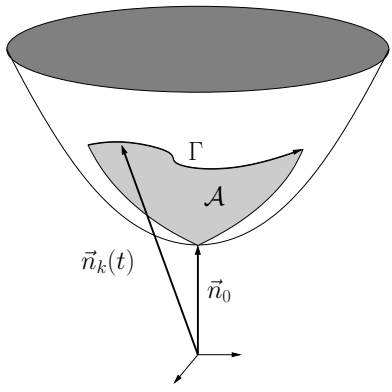}}
               \caption{The vector $\vec{n}_k$ moving on the hyperboloid.}\label{fig:mafigure2}
            \end{center}
         \end{figure}

 From (\ref{Ham:final}) one then finds
 \begin{equation}
 H_{kk+1}(t_s)\equiv \bra{\vec{n}_s} \, H_{k\,k+1}\, \ket{\vec{n}_s}=
  \log\left(1-\frac{\left(\vec{n}_{k,s}-\vec{n}_{k+1,s}\right)^2}{4}\right)\,
  .
\end{equation} The evolution operator is then represented as a product over
small intervals of time $\delta t=\Delta T/ M$. Finally we insert
the expression of unity \eqref{res-un} and then take the limit
$\delta t\to 0$
\begin{eqnarray}\label{ev-split}
 Z
  &=&\tr\e^{-\ii
\Delta T \,\tilde{\lambda}\, H}=\tr\int\,\prod_{s=1}^M   d^2 n_s
\bra{\vec{n}_{s-1}} \e^{-\ii\, \delta t\,\tilde{\lambda}\, H}
  \ket{\vec{n}_s}\nn\\
  &=&\int[\dd n]\e^{\ii S(\vec{n})},\qquad [\dd n]\equiv\prod_s\dd^2
 n_s~.
\end{eqnarray}
 To the leading order in $\delta t$, one finds
\begin{eqnarray}\label{hwz} \bra{\vec{n}_{s-1}} \e^{-\ii\,\tilde{\lambda}\, \delta t\,
H} \ket{\vec{n}_s} &=& \bra{\vec{n}_{s-1}}\vec{n}_s\rangle \, (1-
\ii\,\tilde{\lambda} \,\delta t\, {\bra{\vec{n}_{s-1}} H
\ket{\vec{n}_s}\over
 \bra{\vec{n}_{s-1}}\vec{n}_s\rangle } +\ldots)\nn\\
&=&1+\ii\, \delta t\, \sum_k \,\left[\frac12\left(\cosh\rho_{s,k}-1\right)
\dot{\phi}_{s,k}
-\tilde{\lambda}\, \bra{\vec{n}_s} H_{k\,k+1}\ket{\vec{n}_s}\right]+\ldots \nn\\
&\approx& {\rm exp}\left(\ii\, \delta
t\,\sum_k\,\left[\frac12\left(\cosh\rho_{k}-1\right)
\dot{\phi}_{k}-\tilde{\lambda} \,H_{kk+1} \right](t_s)\right)~,
\end{eqnarray} where we have used (\ref{scalar}) to rewrite
$$
\bra{\vec{n}_{s-1}}\vec{n}_s\rangle\approx 1+\frac{\ii\,\delta
t}{2} \,\sum_k \, \left(\cosh\rho_{k}-1\right) \dot{\phi}_{k}
$$
with $\vec{n}_s=\vec{n}_{s-1}+\delta t \,\dot{\vec{n}}_s+\ldots$~.

Taking the limit $\delta t \to 0$, one finds for the action
\begin{equation}\label{act0} S=\int dt \sum_{k=1}^L \, \left[
\ft12\,\dot{\phi}_k \, \left(\cosh\rho_{k}-1\right)
-\tilde{\lambda}\,
H_{kk+1} \right](t)\, .
\end{equation} As in the su(2) case, the unconventional first-derivative term
in \eqref{act0} is a Wess-Zumino like term. Indeed, it can be
written as a boundary term in one higher dimension $y\in [0,1]$.
Introducing a dependence of the coherent state
$\vec{n}_k=\vec{n}_k(t,y)$ on the extra dimension $y$ and using
\begin{eqnarray}
 \vec{n}\cdot(\pd_y \vec{n}\times\pd_t\vec{n})&=& \sinh\rho\,\left(\partial_y \phi \,\partial_t \rho-\partial_t \phi
 \,\partial_y
 \rho\right)\nn\\
 &=& \partial_{t}\left[ (\cosh \rho -1) \partial_{y} \phi\right]-\partial_{y}\left[ (\cosh \rho -1) \partial_{t} \phi\right]
 \end{eqnarray}
one has for the Wess-Zumino term\footnote{Here and below
$\vec{a}\cdot (\vec{b}\times \vec{c})=\epsilon_{\alpha \beta
\gamma} a^\alpha b^\beta c^\gamma$, $n\cdot m=n^\alpha m_\alpha$
with indices raising and lowering with the metric $\eta_{\alpha
\beta}=(+--)$ .}:
\begin{equation}\label{wz}
  S_{WZ}\equiv \ft12\,\int dt \sum_{k=1}^L \,
\dot{\phi}_k \, \left(\cosh\rho_{k}-1\right)=
   -\ft12 \,\sum_k\int\dd t\,\int_0^1\dd y\,
 \vec{n}_k\cdot (\pd_y\vec{n}_k\times\pd_t\vec{n}_k).
\end{equation}
The two sides in eq. (\ref{wz}) agree for the boundary condition
choice
$$\vec{n}_k(t,0)\equiv\vec{n}_k(t)~,~~~\vec{n}_k(t,1)\equiv
\vec{n}_0 \qquad \pd_y\phi|_{t=\Delta T}=\pd_y\phi|_{t=0}=0~.$$
 For each site, the
double integral (\ref{wz}) is nothing more than the spin $j=\ft12$
times the area $\cal{A}$ between the trajectory of $\vec{n}$ and
the ``north pole'' $\vec{n}_0=(1,0,0)$ (see Figure
(\ref{fig:mafigure2})).
 In addition, $S_{WZ}$ generates the right Poisson brackets
$$
 \{ n_\alpha , n_\beta \}=\epsilon_{\alpha\beta} {}^\gamma \,n_\gamma
 $$
which reproduce the sl(2) commutation relations upon quantization.

The action \eqref{act0} with the two-site Hamiltonian defined as
in \eqref{Ham:final} looks like a discretization of a hyperboloid
sigma model. The correspondence can be made precise by going to
the continuum limit $a={1\over L}\to 0$
 $$
\vec{n}_k\to \vec{n}(\sigma)\big|_{\sigma=ka} \qquad
\sum_{k=1}^L\to {1\over a} \int_0^1 d\sigma  \, .
$$
One finds
\begin{equation}\label{hamcon}
 \tilde{\lambda}\,\sum_k H_{kk+1}(t)\to -{ \tilde{\lambda}\over 4 a} \int d\sigma \, a^2\,(\partial_\sigma \vec{n})^2
 ={ \tilde{\lambda}\over
4 L} \int d\sigma
 \left[(\partial_\sigma \rho)^2+\sinh^2\rho\,(\partial_\sigma
\phi)^2\right] \end{equation} leading to
\begin{equation}
 S= {L\over 2} \int d\sigma d\tau \left( (\cosh\rho-1)\,
\dot{\phi}-{\tilde{\lambda}\over 2\,L^2}
 \left[(\partial_\sigma \rho)^2+\sinh^2\rho\,(\partial_\sigma
\phi)^2\right]\right)\, .\label{Sspin} \end{equation}

 The same result will be found below by considering semiclassical strings
 spinning on $AdS_5\times S^5$. A coherent spin chain state will be specified by
 its spin
\begin{equation} S_z =\sum_{k=1}^L\, \bra{\vec{n}_k} J_0  \ket{\vec{n}_k}\to
{L\over 2}\int \dd \sigma \, \cosh\rho~~. \label{spin}\end{equation}

\section{Spinning strings on $AdS_5\times S^5$}

 Here we describe the string duals of excitations in the sl(2) spin chain
 system. We follow closely \cite{Kruczenski:2003gt}
 where analogous results were found for strings on $S^5$ dual to
 su(2) spin states. Here sl(2) coherent
 states will be associated to classical solutions of the string equations
with non-trivial angular velocities both in $S^5$ and $AdS_5$. We
refer the reader to \cite{Stefanski:2004cw} for an independent
study of these solutions and \cite{Tseytlin:2003ii}
 for a review and a list of references on spinning strings on $AdS_5\times S^5$.

 The bosonic part of Polyakov action describing a string moving on
 $AdS_5\times  S^5$ can be written as
 \begin{eqnarray}
 S={R^2\over 4 \pi \alpha^\prime} \int g_{MN} (\partial_\tau X^M
\partial_\tau
 X^N-\partial_\sigma X^M \partial_\sigma X^N)\label{action}
 \end{eqnarray}
 with
 $$
 ds^2 = g_{MN} dX^M dX^N=ds^2_{AdS_5}+ds^2_{S^5}
 $$
 and
\begin{eqnarray}
 ds^2_{AdS_5}&=&d\rho_0^2-\cosh^2 \rho_0\, dt^2+\sinh^2 \rho_0\,
 (d\theta^2+\cos^2  \theta \,d\phi_1^2+\sin^2 \theta\, d\phi_2^2)\nn\\
 ds^2_{S^5}&=&d\gamma^2+\cos^2 \gamma\, d\varphi_3^2+\sin^2
 \gamma\,
 (d \psi^2+\cos^2  \psi\, d\varphi_1^2+\sin^2 \psi
 \,d\varphi_2^2)\label{metric}
 \end{eqnarray}
 String states are classified by charges with respect to the Cartan
generators $S_{1,2,3}$ and $J_{1,2,3}$ of the SO(4,2) and SO(6)
isometry groups of $AdS_5$ and $S^5$ respectively. In the
spacetime theory they correspond to shifts of $\phi_{1,2,3}$ and
$\varphi_{1,2,3}$ coordinates respectively
\begin{equation}
 S_{1,2}=\partial_{\phi_{1,2}}, \qquad E=\partial_t \qquad
 J_{1,2,3}=\partial_{\varphi_{1,2,3}}\, .
\end{equation}
We are interested in string duals of sl(2) SYM operators made out
of a single scalar $\phi$ and its derivatives. To the direction
$\phi$ we associate a circle on $S^5$ parametrized by $\varphi_3$
while derivatives correspond to excitations along a circle  in
$AdS_5$ parametrized by $\phi_1$.  The remaining excitations will
be turned off. The other choices of the sl(2) sector are related
to this one via SO(4,2)$\times$ SO(6) rotations.

  More precisely we look for solutions of the string equations
  following from (\ref{action}) with
\begin{eqnarray} &&\gamma=\theta=0\qquad \rho=\rho(\sigma,\tau)\qquad
\phi_1=\phi_1(\sigma,\tau)\qquad
\varphi_3=\varphi_3(\sigma,\tau)\quad t=\kappa \tau\, .
\end{eqnarray} In addition, the solutions must satisfy the Virasoro constraints \begin{eqnarray} &&g_{MN} \,\partial_\tau
X^M
\partial_\sigma
 X^N=0\nn\\
&&g_{MN}\, (\partial_\tau X^M  \partial_\tau
 X^N+\partial_\sigma X^M \partial_\sigma X^N)=0\, .\label{vir}
\end{eqnarray}
 As in \cite{Kruczenski:2003gt} we consider the limit $\kappa\to \infty$.
 To this end it is convenient to rewrite the metric in
coordinates where $g_{tt}=0$. This can be done by the following
change of variables: \begin{equation}
 \phi_1= \phi+t \qquad \varphi_3=\varphi+t\qquad \rho_0=\ft12
 \rho\, .
\end{equation}
 The metric in the new variables reads (at $\gamma=\theta=0$)
 \begin{equation}\label{metricnew}
  ds^2=\ft14\, {d\rho}^{2}+\ft12(\cosh\rho-1)\,
 {d\phi}^{2} +d\varphi^2+ dt \,\left[
2\,d\varphi + (\cosh \rho-1) \, d \phi \right]\, . \end{equation}
 Then we consider the limit
\begin{equation}
 \kappa\to \infty \qquad {\rm with}~~~\kappa\, \partial_{\tau} X^{M\neq t}~{\rm
fixed}\qquad  X^M=\phi,\varphi,\rho\, .
 \label{limit}
\end{equation}
 Notice that in the original variables the limit
 corresponds to $\partial_\tau \varphi_3\approx \partial_\tau \phi_1\approx k>>1$, i.e.
the string spins fast on $S^1_{\varphi_3}\in S^5$ and $S^1_\phi\in
AdS_5$. This is expected from holography since states in the sl(2)
sector carry charges $J_3\in$ SO(6) and $E,S_1\in$ SO(4,2).

 To the leading order in $\kappa$, the first of the Virasoro
constraints in (\ref{vir}) reads
\begin{equation} \kappa\, \left[2\,\partial_\sigma
\varphi+(\cosh\rho-1)\,\partial_\sigma \phi\right]=0\, ,
  \label{vir1}\end{equation} which can be used to
solve $\partial_\sigma \varphi $ in favor of $\partial_\sigma
\phi$.

Evaluating (\ref{action}) and using (\ref{metricnew},\ref{vir1}),
one finds (to the leading order in $\kappa$)
\begin{eqnarray}
   S &=&{R^2 \over 4 \pi
   \alpha^\prime} \int d\sigma d\tau \left(\kappa\,\left[(\cosh \rho-1) \,
   \partial_\tau \phi+2\,\partial_\tau \varphi\right] -\ft14
   \,\left[(\partial_\sigma\rho)^2+  \sinh^2
   \rho\,  (\partial_\sigma\phi)^2\right]\right) \, .\nn
\end{eqnarray}
  Finally identifying
\begin{equation}
L={R^2 \kappa \over 2 \pi \alpha^\prime}\qquad
\tilde{\lambda}={R^4\over 8\, \pi^2 \alpha^{\prime 2}}
\end{equation}
and changing to variable $t=k\tau$ one finds\footnote{Notice that
both integrands go like ${1\over \kappa^2}$ in the limit
(\ref{limit}) and therefore the action is finite.}
\begin{eqnarray}
 S  &=&{L\over 2}\, \int d\sigma dt \left( (\cosh \rho-1) \,
   \partial_t \phi+2\,\partial_t \varphi -{\tilde{\lambda}\over 2L^2} \left[
   (\partial_\sigma\rho)^2+  \sinh^2 \rho\,
   (\partial_\sigma\phi)^2\right]\right)\, .\label{act}
\end{eqnarray}
Besides the $\varphi$-dependent total derivative term, the result
(\ref{act}) is in perfect agreement with the spin chain sigma
model result (\ref{Sspin}).
 It can be compared with the result in \cite{Kruczenski:2003gt}
corresponding to strings spinning on $S^5$. The two results differ
(up to total derivative terms) for the replacing of trigonometric
with hyperbolic functions.

\section{Classical solutions}

 The classical string solutions following from the string/spin chain
action (\ref{act}) can be found via analytic continuation from
those in \cite{Kruczenski:2003gt}. In this section we sketch the
results and refer the reader to \cite{Kruczenski:2003gt} for
details.
 In addition we will work out in detail classical solutions
 associated to string sitting at the bottom end of the hyperboloid
 where BMN frequencies and energies are reproduced.

The classical equations of motions are
\begin{eqnarray}
 \sinh \rho
 \,\partial_t\rho-{\tilde{\lambda} \over L^2}\, \partial_\sigma(\partial_\sigma \phi
 \sinh^2 \rho)=0\nn\\ \sinh \rho \,\partial_t \phi+{\tilde{\lambda} \over  L^2}\,\left[
 \partial^2_\sigma \rho- \ft12\sinh(2\rho) (\partial_\sigma
 \phi)^2\right]=0\, .
\label{eom}
\end{eqnarray}
The simplest solutions of eqs. (\ref{eom}) are found by taking
$\partial_\sigma \phi=0$. The two equations then imply
$\partial_\tau \rho=0$ and $\partial_\tau^2 \phi=0$ i.e.
$\rho=\rho(\sigma)$ and $\partial_\tau \phi=\omega$ respectively.
The solution then represents a string with profile $\rho(\sigma)$,
rotating at constant velocity $\omega$ on a circle inside $AdS_5$.
The profile $\rho(\sigma)$ is determined by the second equation in
(\ref{eom})
\begin{equation}\label{eom2}
  \omega\sinh \rho \,+{\tilde{\lambda} \over  L^2}\, \partial_\sigma^2
  \rho=0
\end{equation}
solved by
\begin{equation}
 \partial_\sigma \rho=\pm \sqrt{
 a-b \cosh \rho } \qquad a={\rm const}\quad b={2\, L^2\over \tilde{\lambda}}\,\omega
\end{equation}
with $a$ denoting an integration constant. Solutions exist only
for $a>b$. In this case one can take for $\rho(\sigma)$ an
oscillating solution between $\pm \rho_{\rm max}$ with $\rho_{\rm
max}={\rm Arccosh}\,{a\over b}$. \footnote{Notice that in
coordinates where $\phi\in [0,2\pi]$ the points $(\rho,\phi)$ and
$(-\rho,\phi+\pi)$ are identified.}

 The energy ${\cal E}$ and spin $S_z$ of the string/spin chain state are given by
 (\ref{hamcon},\ref{spin})
\begin{eqnarray}
S_{z}&=& {L\over 2} \int_0^1 \cosh\rho\, \dd\sigma=2\,
L\,\int_0^{\rho_{\rm max}}{\cosh\rho\over \sqrt{a-b
\cosh\rho}} \,\dd\rho\nn\\
&=&2 L\,\sqrt{2\over b} \left[2\, E(x)-K(x) \right]\nn\\
 {\cal E}&=& {\tilde{\lambda} \over 4L} \,\int_0^1 \dd\sigma
\,(\partial_\sigma \rho)^2= {\tilde{\lambda} \over L}
\,\int_0^{\rho_{\rm max}}\sqrt{a-b \cosh\rho}~\dd\rho \nn\\
&=&-2 \sqrt{2\,b}\,{\tilde{\lambda}\over L }
\left[E(x)-(1-x) K(x)\right]~,\label{es}
\end{eqnarray}
in terms of the elliptic integrals
\begin{eqnarray} K(x)&=&\sqrt{b\over 2}\,\int_0^{\rho_{\rm max}} {1\over \sqrt{a-b \cosh\rho}}
\,d\rho  \qquad x={b-a\over 2\,b}\nn\\[3mm]
E(x)&=&\frac12\,\sqrt{b\over 2}\int_0^{\rho_{\rm max}} \frac{1+\cosh\rho}{\sqrt{a-b
\cosh\rho}}
\,d\rho~~.
\end{eqnarray}
 Eqs. (\ref{es}) has to be evaluated on $x$, determined by
\begin{equation} \label{idx}
1=\int_0^1 \dd \sigma=4 \int_0^{\rho_{\rm max}} {1\over \sqrt{a-b
\cosh\rho}} \,d\rho=4\sqrt{2\over b}\, K(x)\, .
\end{equation}
  Eqs. (\ref{es},\ref{idx}) describe a
 string/spin chain solution of
length $L$, total spin $S_z$ and anomalous dimension/energy ${\cal
E}$.

 It is instructive to consider solutions localized near the
 bottom end of the hyperboloid with $\rho_{\rm max}\ll 1$.
In this limit one can consider the linearized version of
\eqref{eom}
\begin{eqnarray}
 \rho
 \,\partial_t\rho-{\tilde{\lambda} \over L^2}\, \partial_\sigma(\partial_\sigma \phi
 \rho^2)=0\nn\\
 \rho \,\partial_t \phi+{\tilde{\lambda} \over L^2}\,\left[
 \partial^2_\sigma \rho- \rho (\partial_\sigma \phi)^2\right]=0.
\label{eom-lin}
\end{eqnarray}
 It is interesting to note that similar equations appear also in
the case of su(2) sector with $\rho\to \theta,\phi\to \varphi$.
This is the clear case because at low scales $\rho_{\rm max}$ the
string does not distinguish the hyperboloid from the sphere.

As before, Eq. \eqref{eom-lin} can be easily solved by taking
$\partial_\sigma \phi=0$.
  The system reduces to the harmonic oscillator
equation \begin{eqnarray}
  &&\partial_\sigma^2 \rho+ \nu^2 \rho=0,\quad \nu^2\equiv \frac{\omega
  L^2}{\tilde{\lambda}}\nn\\
  &&\quad \partial_t \phi=\omega \qquad \partial_t \rho=0\, .
\end{eqnarray} For $\omega>0$ this leads
to the harmonic solution
\[
  \rho=\rho_{\rm max}\cos \nu(\sigma-\sigma_0)\, .
\]
Requirement of periodicity in $\sigma$:
$\rho(\sigma+1)=\rho(\sigma)$ leads to quantization condition for
$\nu$
\begin{equation}\label{qOmega}
  \nu_n=2\pi n, \qquad n\in \mathbb{Z}\, .
\end{equation}
For the above solution is not difficult to compute the integrals
of motion of interest:
\begin{align}\label{sz}
  S_z&=\frac{L}{2}\int_0^1\dd\sigma\,(1+\ft12 \,\rho^2+\ldots)=\frac{L}{2}+\frac{L}{2}\frac{\rho^2_{\rm
  max}}{4}+\ldots \\ \label{e}
  \mathcal{E}&=\frac{\tilde{\lambda}}{4 L}\, \int_0^1\dd\sigma\,(\pd_\sigma\rho)^2=
\frac{\tilde{\lambda}\,\rho_{\rm max}^2\, \nu_n^2}{8
L}=\frac{\rho_{\rm max}^2\, \omega_n L}{8}\, .
\end{align}

From \eqref{sz} one can express $\rho_{\rm max}$ in terms of the
spin $s=S_z-L/2$\footnote{Notice that in our conventions vacuum
corresponds to the state with minimal total spin $L/2$. }. Then
the energy/anomalous dimension of a string/spin chain state at
level $n$ and spin $s$ is given by
\begin{equation}\label{E-vs-sz}
  \mathcal{E}_{s,n}=4\pi^2n^2\, s\,  {\tilde{\lambda} \over L^2} \, .
\end{equation}
Formula (\ref{E-vs-sz}) is in perfect agreement with the
expectations coming from BMN analysis.
 It would be interesting to compare this equation to the lowest energy
levels in sl(2) spin chain produced by Bethe ansatz. Let us note
also that the same equation for energy is valid for the su(2)
sector as well.

\section{Conclusions}

In this paper we derived a coherent state representation for the
Hamiltonian of a spin chain with symmetry group sl(2). Coherent
states for sl(2) are in one-to-one correspondence with points on a
hyperboloid and carry a natural coset structure sl(2)/u(1). The
result for the sl(2) Hamiltonian can be compared with its su(2)
analog
\begin{eqnarray}
 H_{su(2)} &=&\ft12 \sum_k (\vec{n}_k-\vec{n}_{k+1})^2\nn\\
 H_{sl(2)} &=& \sum_k \,\log\left(1-\frac{\left(\vec{n}_k-\vec{n}_{k+1}\right)^2}{4}\right)
\end{eqnarray}
 with $\vec{n}$ living on $S^2$ for su(2) and on a two-dimensional
 hyperboloid for sl(2).
The two results are similar\footnote{As expected they are related
to each other via an analytic continuation.} but there are
important differences. First, by being sl(2) non-compact,
representations are infinite-dimensional and therefore coherent
states involve a linear combination of an infinite number of
states. Second, in contrast with the su(2) case, the sl(2)
Hamiltonian has a non-polynomial character.

In the continuum limit (where the number of chain sites gets
large), both Hamiltonians reduce to that of string sigma models on
the sphere/hyperboloid spaces. We derive a path integral
formulation of the sl(2) spin chain dynamics. As in the su(2) the
resulting Lagrangian contains a Wess--Zumino term, which ensures
the right commutation relations between the hyperboloid
coordinates upon quantization.

Spin chain coherent states are identified with strings spinning
fast on a torus $S_\phi^1\times S_\varphi^1$ with $S_\phi^1\in$
$AdS_5$ and $S_\varphi^1\in S^5$. The spin chain and string sigma
models are shown to be in perfect agreement in consistency with
the results of \cite{Stefanski:2004cw}. In addition we consider
classical solutions of the strig/spin chain sigma model actions.

As expected, the results are often related to those in su(2) via
an analytic continuation.

There are several interesting directions to go on. One can extend
the analysis here either for higher number of loops, or by
considering the first non-planar corrections where the hamiltonian
(although no longer integrable) is known. In addition, it would
nice to generalize our results to the case of the psu(2,2$|$4)
spin chain describing ${\cal N}=4$ SYM. At this stage the
generalization looks conceptually straightforward although
technically involved.

The sl(2) symmetry is common to any (asymmptotically) free gauge
theory since it is part of the conformal group in arbitrary
dimensions, for instance see \cite{Lipatov:1993yb,Faddeev:1994zg}
for studies in the framework of QCD and \cite{Belitsky:2004cz} for
a review of recent development and a complete list of references.
We believe that our results here can be helpful in further studies
along these lines. It would be nice to understand whether the
results here generalize to gauge theories in $d\neq 4$ and their
string dual on warped AdS spaces
\cite{Itzhaki:1998dd,Gherghetta:2001iv,Morales:2002ys,Morales:2002uh}.

\subsection*{Acknowledgements}

We thank A. Tseytlin for pointing out our attention to the recent
paper \cite{Stefanski:2004cw} where part of the results here were
independently found.
 We thank E. Orazi and H. Samtleben for
discussions and G. Arutyunov and A. Belitsky for correspondence.
This work was partially supported by NATO Collaborative Linkage
Grant PST.CLG. 97938, INTAS-00-00254 grant, INTAS-00-00262, RF
Presidential grants MD-252.2003.02, NS-1252.2003.2, INTAS grant
03-51-6346, RFBR-DFG grant 436 RYS 113/669/0-2, RFBR grant
03-02-16193 and the European Community's Human Potential Programme
under contract HPRN-CT-2000-00131.

\begin{appendix}

\section{Coherent states}
\label{acs}
 We prove here  important relations concerning the coherent
 states defined in Section 2. These are defined with the infinite
 series
$$\ket{\vec{n}(\rho,\phi)}=\sum_{m=0}^{\infty}c_{j,m}(\rho,\phi)\ket{j,j+m},\qquad
  c_{j,m}(\rho,\phi)=\frac{1}{\cosh(\ft{\rho}{2})^{2j}}
\left(\frac{\Gamma(m+2j)}{m!\,\Gamma(2j)}\right)^{\frac12}\,\e^{\ii\,m\,\phi}\,\tanh(\ft{\rho}{2})^{m}.$$
 Using
$$\sum_{m=0}^{\infty}\frac{\Gamma(m+2j)}{m!\,\Gamma(2j)}\,x^m=\left(1-x\right)^{-2j}~,$$
it is easy to prove that
$$\left\{\begin{array}{lll}
\dst\left<\vec{n}|\vec{n}\right>&=\sum_{m=0}^{\infty}\bar{c}_{j,m}
c_{j,m}
=\cosh(\frac{\rho}{2})^{-2j}\,(1-\tanh^{2}(\frac{\rho}{2}))^{-2j}&=1\\[3mm]
\dst \left<\vec{n}|J_0|\vec{n}\right>
&=\sum_{m=0}^{\infty}\bar{c}_{j,m} c_{j,m}(m+j)
=j\,\cosh(\frac{\rho}{2})^{-2j}\,\frac{1+\tanh^{2}(\frac{\rho}{2})}{(1-\tanh^{2}(\frac{\rho}{2}))^{2j+1}}
&=j\,\cosh\rho\\[3mm]
\dst
\left<\vec{n}|J_+|\vec{n}\right>&=\sum_{m=1}^{\infty}\bar{c}_{j,m}
c_{j,m-1}\sqrt{m(m-1+2j)}
=j\,\frac{\e^{-i\phi}}{\cosh(\frac{\rho}{2})^{2j}}\frac{2\,\tanh(\frac{\rho}{2})}{(1-\tanh(\frac{\rho}{2}))^{2j+1}}
&=j\,\e^{-i\phi}\sinh\rho\\[3mm]
\dst
\left<\vec{n}|J_-|\vec{n}\right>&=\sum_{m=0}^{\infty}\bar{c}_{j,m}
c_{j,m+1}\sqrt{(m+1)(m+2j)}
=j\,\frac{\e^{i\phi}}{\cosh(\frac{\rho}{2})^{2j}}\frac{2\,\tanh(\frac{\rho}{2})}{(1-\tanh(\frac{\rho}{2}))^{2j+1}}
&=j\,\e^{i\phi}\sinh\rho
\end{array}\right.$$

In order to prove the resolution of unity (\ref{res-un}), we act
with its \emph {l.h.s} on an arbitrary state $\left|j,m \right>$:
$$\begin{array}{rl}
\dst\int\dd \rho\,\dd \phi\sinh \rho\,
  \ket{\vec{n}}\bra{\vec{n~}}\left.j,j+m \right> &={\dst\sum_{k=0}^{\infty}\int}\dd \rho\,\dd \phi
 \sinh \rho\, \frac{\tanh(\frac{\rho}{2})^{k+m}}{\cosh(\frac{\rho}{2})^{4j}}\sqrt{\frac{\Gamma(m+2j)\Gamma(k+2j)}
 {k!\,m!\,\Gamma(2j)^2}}\,\e^{\ii(k-m)\phi}\ket{j,j+k}\\[5mm]
  &\dst=2\,\pi\,\frac{\Gamma(m+2j)}{m!\,\Gamma(2j)}\int_0^\infty\dd \rho\,
 \sinh \rho\,
 \frac{\tanh(\frac{\rho}{2})^{2m}}{\cosh(\frac{\rho}{2})^{4j}}\ket{j,j+m}\\[5mm]
  &\dst=8\,\pi\frac{\Gamma(m+2j)}{m!\,\Gamma(2j)}\int_0^\infty\dd \tau\,
 \frac{(\sinh\tau)^{2m+1}}{(\cosh\tau)^{4j+2m-1}}\ket{j,j+m}~.
  \end{array}$$
As
$$\int_0^\infty\dd
\rho\,\frac{(\sinh\rho)^{A}}{(\cosh\rho)^{B}}=\frac{\Gamma(\frac{A+1}{2})\Gamma(\frac{B-A}{2})}{2\,\Gamma(\frac{B+1}{2})}\quad\quad
\hbox{ for }A<B~,$$
one ends with
$${2j-1\over 4 \pi}\,\int\dd \rho\,\dd \phi\sinh \rho\,
  \ket{\vec{n}}\bra{\vec{n~}}\left.j,j+m \right> = (2j-1)\frac{\Gamma(2j-1)}{\Gamma(2j)}\ket{j,j+m}=
  \ket{j,j+m}\quad\quad
 $$
 for any $j>1/2$. Notice that also the limit $j\to 1/2$ is well defined.
 This proves the identity (\ref{res-un}).

\section{Hamiltonian in the CS representation}
\label{aham}

We restrict ourselves here to the $j=\ft12$ case. The two-site
Hamiltonian can then be rewritten as
$$H_{k_1 k_2}\ket{m_1 m_2} = (h(m_1)+h(m_2))\ket{m_1 m_2}-\sum_{l=1}^{m_1}\frac{1}{l}\ket{m_1-l,m_2+l}
  -\sum_{l=1}^{m_2}\frac{1}{l}\ket{m_1+l,m_2-l}~.
$$
Acting on coherent states, it gives
$$\begin{array}{rl}
 H_{k_1 k_2}\left|\vec{n}_1,\vec{n}_2\right> &=\dst
\sum_{m_1=0}^{\infty}\sum_{m_2=0}^{\infty}c_{m_1}(\rho_1,\phi_1)\,c_{m_2}(\rho_2,\phi_2)\,\Bigg[(h(m_1)+h(m_2))\ket{m_1,m_2}~~~~~~~~~~\\
&\dst\hfill -\sum_{l=1}^{m_2}\frac{1}{l}\left|m_1+l,m_2-l\right>
-\sum_{l=1}^{m_1}\frac{1}{l}\left|m_1-l,m_2+l\right>\Bigg]~.
\end{array}$$
The sums over $l$ can be extended to infinity using the relation
$$\sum_{m=1}^{\infty}\sum_{l=1}^{m}=\sum_{M=m-l=0}^{\infty}~~\sum_{l=1}^{\infty}$$
  We get ~:
$$\begin{array}{ll}
\displaystyle
\left<\vec{n}_1\vec{n}_2\right|H_{12}\left|\vec{n}_1\vec{n}_2\right>=
&\displaystyle
\frac{1}{\cosh(\frac{\rho_1}{2})^{2}\cosh(\frac{\rho_2}{2})^{2}}\sum_{m_1=0}^{\infty}\sum_{m_2=0}^{\infty}
\tanh(\frac{\rho_1}{2})^{2m_1}\tanh(\frac{\rho_2}{2})^{2m_2}(h(m_1)+h(m_2))\\
&\displaystyle
-\frac{1}{\cosh(\frac{\rho_1}{2})^{2}\cosh(\frac{\rho_2}{2})^{2}}\sum_{m_1=0}^{\infty}\sum_{M_2=0}^{\infty}
\sum_{l=1}^{\infty}
 \e^{-i\,l\,(\phi_1-\phi_2)}\tanh(\frac{\rho_1}{2})^{2m_1+l}\tanh(\frac{\rho_2}{2})^{2M_2+l}\\
&\displaystyle
-\frac{1}{\cosh(\frac{\rho_1}{2})^{2}\cosh(\frac{\rho_2}{2})^{2}}\sum_{M_1=0}^{\infty}\sum_{m_2=0}^{\infty}
\sum_{l=1}^{\infty}
 \e^{i\,l\,(\phi_1-\phi_2)}\tanh(\frac{\rho_1}{2})^{2M_1+l}\tanh(\frac{\rho_2}{2})^{2m_2+l}
\end{array}$$
The first term sums
to
$$\log\left(\cosh(\frac{\rho_1}{2})^{2}\cosh(\frac{\rho_2}{2})^{2}\right)~,$$
the second and third terms to
$$\log\left(1-\e^{\mp
i(\phi_1-\phi_2)}\tanh(\frac{\rho_1}{2})\tanh(\frac{\rho_2}{2})\right)~.$$
Adding the three logarithms yields finally
$$\begin{array}{ll}
\displaystyle\left<\vec{n}_1,\vec{n}_2\right|H_{12}\left|\vec{n}_1,\vec{n}_2\right>&\displaystyle=\log\frac12\left(1+
\cosh\rho_1\,\cosh\rho_2-\cos(\phi_1-\phi_2)\sinh\rho_1\,\sinh\rho_2\right)\\[5mm]
&\displaystyle=\log\frac12\left(1+\vec{n}_1.\vec{n}_2\right)\\[5mm]
&\displaystyle=\log\left(1-\frac{\left(\vec{n}_1-\vec{n}_2\right)^2}{4}\right)
\end{array}$$
where as before the product is defined via
$$
 \vec{n}.\vec{m}=n_0 m_0-n_1 m_1-n_2 m_2~.
 $$

\end{appendix}

\providecommand{\href}[2]{#2}\begingroup\raggedright\endgroup

\end{document}